\documentclass[aps,prev,twocolumn,preprintnumbers,floatfix,nofootinbib]{revtex4-1}
\pdfoutput=1

\newcommand{\be}{\begin{equation}}
\newcommand{\ee}{\end{equation}}
\newcommand{\bea}{\begin{eqnarray}}
\newcommand{\eea}{\end{eqnarray}}

\usepackage{algorithm}
\usepackage[noend]{algpseudocode}

\usepackage{mathrsfs}
\usepackage{amsmath, amsthm, amssymb}
\usepackage{color}
\usepackage{epstopdf}
\usepackage[pdftex]{graphicx}
\usepackage{amssymb}
\usepackage{verbatim}
\usepackage{hyperref}
\usepackage{enumerate}

\newcommand{\kcite}[1]{Ref.~\cite{#1}}

\newcommand{\keq}[1]{Eqn.~(\ref{#1})}
\newcommand{\kfig}[1]{Fig.~\ref{#1}}

\newcommand{\kap}[1]{App.~\ref{#1}}
\newcommand{\ksec}[1]{Sec.~\ref{#1}}

\begin{document}

\preprint{KCL-PH-TH/2018-29}

\title{Formation of Relativistic Axion Stars}

\author{James Y. Widdicombe$^{a}$}
\email{j.y.widdicombe@gmail.com}
\author{Thomas Helfer$^{a}$}
\email{thomashelfer@live.de}
\author{David J. E. Marsh$^{b}$}
\email{david.marsh@uni-goettingen.de}
\author{Eugene A. Lim$^{a}$}
\email{eugene.a.lim@gmail.com}

\vspace{1cm}
\affiliation{
${}^a$ Theoretical Particle Physics and Cosmology Group, Physics Department, Kings College London, Strand, London WC2R 2LS, United Kingdom}
\affiliation{${}^b$Instit{\"u}t f{\"u}r Astrophysik, Georg-August Universit{\"a}t, Friedrich-Hund-Platz 1, D-37077 G{\"o}ttingen, Germany}

\begin{abstract}

Axions and axion-like particles are compelling candidates for the missing dark matter of the universe. As they undergo gravitational collapse, they can form compact objects such as axion stars or even black holes. In this paper, we study the formation and distribution of such objects. First, we simulate the formation of compact axion stars using numerical relativity with aspherical initial conditions that could represent the final stages of axion dark matter structure formation. We show that the final states of such collapse closely follow the known relationship of initial mass and axion decay constant $f_a$. Second, we demonstrate with a toy model how this information can be used to scan a model density field to predict the number densities and masses of such compact objects. In addition to being detectable by the LIGO/VIRGO gravitational wave interferometer network for axion mass of $10^{-9} < m_a < 10^{-11}$ eV, we show using peak statistics that for $f_a < 0.2M_{pl}$, there exists a ``mass gap'' between the masses of axion stars and black holes formed from collapse.

\end{abstract}

\maketitle

\section{Introduction}

The Advanced Laser Interferometry Gravitational wave Observatory (LIGO) has made historic measurements of gravitational waves (GW) from the binary coalescence of black holes \cite{2016PhRvL.116f1102A} and neutron stars \cite{2017PhRvL.119p1101A}. This paves the way for searches for signals from ``exotic compact objects'' (ECO; see e.g. \cite{Alcubierre:2003sx,2016JCAP...10..001G,Cardoso:2016oxy,Hui:2016ltb,Palenzuela:2006wp,Brito:2015yfh,Hanna:2016uhs}). Axions and axion-like particles \cite{Berezhiani:1989fu,Berezhiani:1989fp,Sakharov1994id,Berezhiani:1992rk,pecceiquinn1977,weinberg1978,wilczek1978,2014JHEP...06..037D,2010ARNPS..60..405J,2006JHEP...06..051S,axiverse,2006JHEP...05..078C,Marsh:2015xka,Cicoli:2012sz} (which we refer to collectively as simply ``axions'') can form such ECO, known as axion stars, which are related to a family of compact scalar field (pseudo)-solitons including Wheeler's ``geons'', boson stars, and oscillatons \cite{1968PhRv..172.1331K,1969PhRv..187.1767R,Liddle:1993ha,1994PhRvL..72.2516S,1991PhRvL..66.1659S}. 

To have a strong GW signal in LIGO, the ECO must have mass and compactness, 
\begin{equation}
\mathcal{C}\equiv\frac{GM_{\star}}{R},
\end{equation}
where $M_\star$ is the mass of the object,  in a particular range \cite{2016JCAP...10..001G}.  Simulations have shown that there are known environments in dark matter halos in which non-relativistic axion stars form \cite{Du:2018zrg,Veltmaat:2018dfz,Schive:2014dra,2016PhRvD..94d3513S,Levkov:2018kau}. Any source of large (possibly primordial) density perturbations, or rapid merging and accretion could potentially grow such stars into the range of mass and compactness accessible to LIGO, and even beyond as they collapse to BH or disperse as novae \cite{2017JCAP...03..055H}. However, there have not been simulations of the final stages of axion star formation in the full relativistic regime and beyond spherical symmetry, which are required to determine the fate of large axion densities.

In the following we simulate the formation of compact axion stars and BH from some pseudo-random initial conditions using full (3+1) dimensional numerical relativity simulations with \textsc{GRChombo} \cite{2015CQGra..32x5011C}. Our results can be used to assess axion star formation given some input realisation of the axion density field. We demonstrate this for a toy model density field, using peak statistics to label compact axion stars and BH in the LIGO frequency band. 

We remain agnostic about the amount of dark matter (DM) that might be contained in compact axion stars, noting only that it must be relatively small, of order a few percent (see e.g. the compilations of primordial BH constraints in \kcite{Josan:2009qn,Carr:2016drx,2017arXiv170102151N}). Given the theoretical uncertainties in formation mechanisms for compact axion stars from axion dark matter, such bounds can easily be consistent with all the dark matter being axions.

We use units $\hbar=c=1$ and $M_{pl}=1/\sqrt{8\pi G}$ is the reduced Planck mass.

\section{Gravitational Waves from ECOs}
\begin{figure}
\center
\includegraphics[width=1\columnwidth]{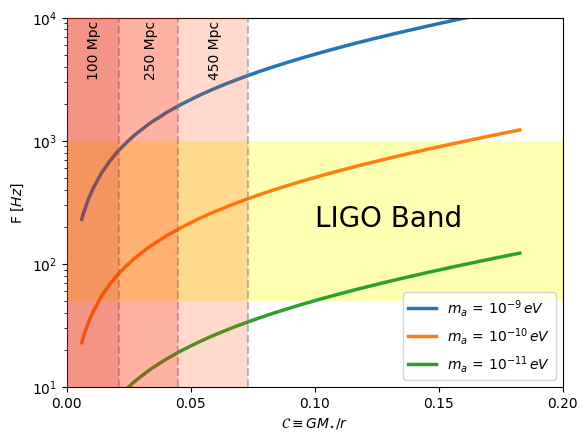}
\caption{LIGO frequency band for axion stars. The frequency is given by the ISCO frequency, Eq.~(\ref{eqn:fcomp}). The compactness $\mathcal{C}(M)$ is for non-interacting oscillatons, which is a good description of stable axion stars. Assigning a luminosity distance to binaries, the minimum compactness is found from the results of Ref.~\cite{2016JCAP...10..001G}. Axion stars detectable by LIGO must have $\mathcal{C}\gtrsim 0.02$ and axion mass $m_a\approx 10^{-10}\text{ eV}$.}
\label{fig:ligo_freq}
\end{figure}

In the following we review the description of GWs from ECOs given in \kcite{2016JCAP...10..001G}. An ECO is described by two parameters, the mass, $M_{\star}$, and the compactness, $\mathcal{C}$, which together determine the frequency and amplitude of GWs produced in a binary inspiral (the merger and ringdown phase contain more information requiring direct simulation). The orbital period, $P$, is related to the total binary mass, $M_{tot}$ and semi-major axis, $l$, by Kepler's third law
\begin{equation}
P^2 = \frac{4 \pi^2 l^3}{GM_{tot}} \, .
\end{equation}
The frequency, $f$, of gravitational wave emission is twice \footnote{As both frequency, $f$, and orbital frequency, $\nu$, contain factors of $2\pi$ we can cancel them for ease} the orbital frequency, $\nu = 1/P$, and hence is given by
\begin{equation}
f = \sqrt{\frac{GM_{tot}}{\pi^2 l^3}} \, .
\end{equation}
The innermost stable circular orbit, ISCO, determines the end of the inspiral phase, and the beginning of the merger phase. For a blackhole binary, the ISCO is given by 
\begin{equation}
R^{\rm ISCO}_{\rm BH} = 6 GM_{tot} \, .
\end{equation}
For an ECO, the ISCO is modified by the variable compactness:
\begin{equation}
R^{\rm ISCO}_{\rm ECO} = \frac{3 GM_{tot}}{\mathcal{C}}\, .
\end{equation}
 Hence the typical frequency, $f^{\rm ISCO}_{\rm ECO}$, of two merging ECO is 
\begin{equation}
\label{eqn:fcomp}
f^{\rm ISCO}_{\rm ECO} = \frac{\mathcal{C}^\frac{3}{2}}{3^\frac{3}{2} \pi G M_{tot}}\, .
\end{equation}

For blackholes, the maximum frequency for gravitational wave emission at the end of the inspiral is given by numerical relativity, and is defined as $f= ( 1 + \Delta) f^{\rm ISCO}_{\rm BH}$. $\Delta$ is a correction term computed in post-Newtonian approximation \cite{Ajith:2009bn}, and is dependant of the mass ratio and spins of the blackholes. In the parameter range where the post-Newtonian approximation is valid, it was found that $\Delta = \mathcal{O}(1)$, although we only expect this to hold for relatively compact objects with ${\cal C}$ within a factor of a few of BH. With this knowledge, for the discussion we deem it adequate to take \keq{eqn:fcomp} as the typical frequency of gravitational wave emission.  

The LIGO noise-power spectral density is minimized between 50 Hz and 1000 Hz. Placing the frequency in Eq.~\ref{eqn:fcomp} in the LIGO band gives the range of $M$ and $\mathcal{C}$. For BH we find the benchmark mass for LIGO of $M_{\star}\approx 10 M_\odot$ stellar mass BH, while for LISA one finds sensitivity to supermassive BHs, $M_{\star}\approx 10^3-10^7 M_\odot$. Ref.~\cite{2016JCAP...10..001G} considered the signal to noise for ECO binary mergers in the LIGO band, and found that for events within a given luminosity distance $D_L$ there is a minimum value of ${\cal C}$ at any given mass to given an event with large signal to noise. 

Fig.~\ref{fig:ligo_freq} shows the results of Ref.~\cite{2016JCAP...10..001G} for the minimum compactness for different luminosity distances together with the $\mathcal{C}(M)$ relation for axion stars determined from spherically symmetric numerical GR~\cite{2017JCAP...03..055H,Helfer:2018vtq}. Axion stars detectable by LIGO must have $\mathcal{C}\gtrsim 0.02$ and axion mass $m_a\approx 10^{-10}\text{ eV}$, implying that the axion stars are of approximately solar mass. 

\section{Axion Star Formation}
\label{sec:initial}

Axion stars giving rise to potential GW inspiral signals in LIGO have high compactness, and are thus relativistic objects. If axion stars can reach such high compactness, they could also surpass their maximum stable mass, entering the unstable region in the ``phase diagram''~\cite{2017JCAP...03..055H,Michel:2018nzt,2017PhRvL.118a1301L}, either collapsing to a BH or dispersing in a nova, depending on the axion ``decay constant'', $f_a$. 

\subsection{Initial Conditions}

We consider an initial state of energy density in the axion field characterised by a single momentum scale, $k_\star$, in a superposition of waves in $(x,y,z)$:
\begin{equation}
\phi = \varphi \left[\cos k_\star x+\cos k_\star y+\cos k_\star z \right]\, .
\label{eqn:initial_phi}
\end{equation}
The waves have initially zero velocity, $\dot{\phi}=0$. We impose periodic boundary conditions on our computational domain, and hence the initial condition is that of a superposition of waves that is not spherically symmetric, but possesses a 6-fold discrete permutation symmetry. This breaks spherical symmetry for the density peak, allowing us to investigate the effects of anisotropy while keeping the parameter space sufficiently small so that we can scan through them with available computational resources. This initial condition represents a locally overdense region dominated by the axion energy density, and hence is decoupled from the Hubble flow.

\begin{figure}
\centering
\includegraphics[width=1\columnwidth]{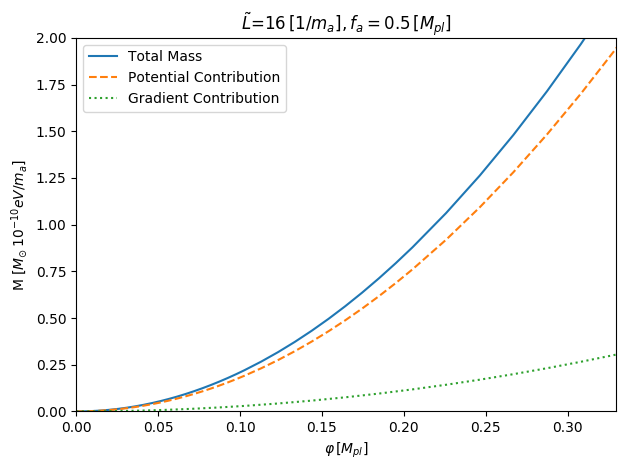}
\caption{Using \keq{eqn:dimlessmass} we calculate the total box mass, $M$, of our initial conditions for a box size of $\tilde{L}=16 \, m_a^{-1}$ and $f_a = 0.5 \, M_{pl}$, as well as the contributions to $M$ of the gradient term, $\frac{1}{2}(\partial_i \phi)^2$, and the potential term, $V(\phi)$. For the $f_a$ simulated it can be seen that our initial mass is dominated by contributions from the potential term. }
\label{fig:initial_condition}
\end{figure}

According to the results of Ref.~\cite{2017JCAP...03..055H} we expect collapse to be governed by two parameters: the total mass, $M$, in a single overdensity (periodic boundary conditions means that this is half the total box mass), and the axion decay constant, $f_a$, defining a ``phase diagram''.\footnote{This diagram has been explained by various arguments in Refs.~\cite{Chavanis:2017loo,Visinelli:2017ooc}. The phase boundaries have been accurately determined using spherically symmetric simulations by \kcite{Michel:2018nzt}. See also the simulations of \kcite{2017PhRvL.118a1301L} who study the regime of low $f_a$ and low curvature leading to axion emission.} The total maximum mass in an overdensity is found by integrating the initial potential energy inside the box, and dividing by two (as two objects will form due to symmetry):
\begin{equation}
\frac{1}{2} \int_V  \rho \sqrt{\det \gamma_{ij}} dV\, , \label{eqn:mass_integral}
\end{equation}
with 
\begin{equation}
\rho = n^\mu n^\nu T_{\mu\nu} \, ,
\end{equation}
where $n^{\mu}$ is the normal to the hypersurface and $\gamma_{ij}$ the 3-D spatial metric. Assuming that the metric is conformally flat as our initial energy density has a small average density, then using \keq{eq:App:rho_initial}
\begin{equation}
M = \frac{1}{2}\int_{-L/2}^{L/2}{\rm d}x~{\rm d}y~{\rm d}z \left (  \frac{1}{2}(\partial_i \phi)^2 + V(\phi) \right )\, , \label{eqn:dimlessmass}
\end{equation}
where $L \equiv 2\pi/k_*$ is the physical size of the  periodic domain.
The axion potential energy is given by
\begin{equation}
V(\phi) = m_a^2f_a^2 \left[1-\cos\left(\frac{\phi}{f_a}\right)\right] \, .
\end{equation}

By a choice of units, the axion mass $m_a$ can be scaled out of all our simulations; units can be easily restored to set the physical mass of the compact objects formed. To achieve this scaling, we chose
\begin{equation}
M = 0.27 \tilde{M}\left(\frac{10^{-10}\text{ eV}}{m_a}\right)M_\odot\, 
\end{equation}
where $M_\odot$ is the solar mass. Meanwhile, for small $\tilde{L}<(\varphi/f_a)$ the gradient term dominates.  \kfig{fig:initial_condition} shows the initial conditions for the smallest $\tilde{L}$ we simulated and for $f_a = 0.5 \, M_{pl}$. All of our numerical simulations had initial conditions where potential energy dominated.

Finally, we can compute the average energy density of the simulation domain  via
\begin{equation}
\bar\rho = \frac{M}{L^3}  \, ,
\end{equation}
and hence the ``local'' Hubble constant $H^2_{\rm local} = (1/3M^2_{pl})\bar\rho$ which is $H_{\rm local} \sim m_a$. We emphasise that this is not related to the ``global'' Hubble constant, since we are simulating a local overdense region. 

Our simulations begin at an arbitrary dimensionless time, and we should ask how this is related to the cosmic time. In our simulations, we are evolving an axion dominated overdense local patch that is much smaller than the current Hubble radius, thus it is assumed that the expansion of the Universe and the presence of any fluctuation in energy density of non-axion components can be neglected. Thus, a fluctuation of any amplitude in our simulations will collapse, and cosmologically we cannot relate this to the collapse threshold for a given redshift.

A perturbation mode of \emph{co-moving} wave number $k$ with frequency $\omega^2(a) = (k/a)^2 + m_a^2$ will begin to evolve when $\omega(a)> H(a)$. Consider a co-moving mode $k_{\rm cm}$ which re-enters the horizon at time $a_{\rm ret}$, $k_{\rm cm} = a_{\rm ret} H(a_{\rm ret})$. Furthermore, if $H(a_{\rm ret})< m_a$, then the mode will collapse only at time $a_{\rm osc} > a_{\rm ret}$ where $H(a_{\rm osc}) \approx m_a$, i.e. when the mode is subhorizon. In our simulations, our box size is set to $\tilde{L} = 2\pi m_a/k_\star$ where $k_\star$ is a \emph{physical scale} and related to the co-moving wave vector $k_{\rm cm}$ by an arbitrary scale factor $a$.

From our choice of dimensionless units, this means that the physical length $k_\star^{-1} = (\tilde{L}/2\pi)m_a^{-1}$. As will be described in section \ref{sect:num_setup}, we take $\tilde{L} ={\cal O}(16\sim128)$ in our simulations, and hence $k_\star < m_a$, which satisfies the condition for subhorizon collapse above.

\subsection{Numerical Simulations}\label{sect:num_setup}

\begin{figure*}
\centering
\includegraphics[width=1\columnwidth]{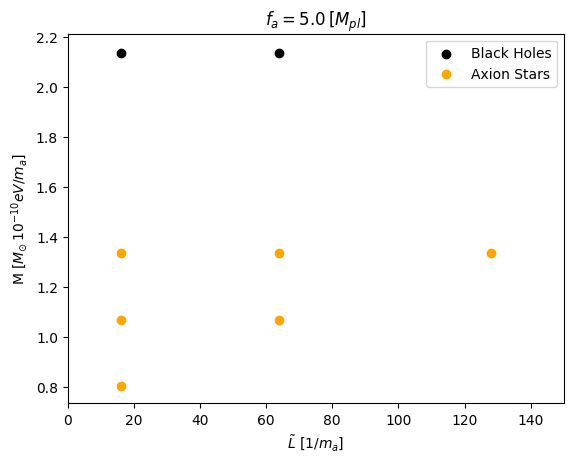}
\includegraphics[width=1\columnwidth]{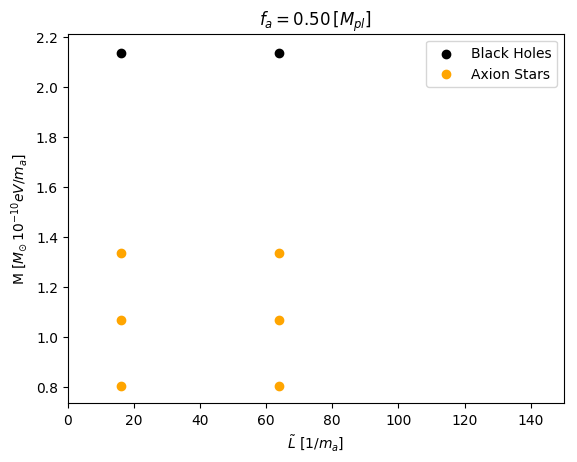}
\caption{These plots are a summary of all numerical simulations performed. Black circles indicate that blackholes where formed from the initial conditions and yellow circles indicate that axion stars were formed. We emphasise that the y-axis labels the \emph{initial total mass} of the simulation initial conditions, not the final mass of the formed objects. No dispersion cases were obtained, and the reason for this is outlined in \ksec{sec:dispersion}. The results presented here mirror that of \kcite{2017JCAP...03..055H,Michel:2018nzt}, so when discussing the likely structure formation we will use the ``phase diagram" constructed there. 
\label{fig:moneyplot}}
\end{figure*}

We simulated collapse of a massive scalar field, $\phi$, with an axion potential in numerical relativity, using $\textsc{GRChombo}$ \cite{Clough:2015sqa}. Details of the numerical scheme can be found in \kap{sect:GRChombo}. We probed a three dimensional ``phase diagram", summarised in \kfig{fig:moneyplot}, to investigate how collapse differed whilst varying initial mass $M$, length scale of the axion waves $\tilde{L}$, and decay constant $f_a$. In the following sections we will explore the different types of structure that can be formed according to the parameters of the ``phase diagram", as well as commenting on the technical limitations that we faced.

To explore the possible ``phase diagram" of initial conditions for axion star collapse, we choose three length scales of the axion waves $\tilde{L}=16,64,128$ $m_a^{-1}$, two decay constants $f_a=5.0, 0.5$ $M_{pl}$ and four initial total box masses $M=2.14,1.34,1.07,0.80$ $M_{\odot} \, (10^{-10} \mathrm{eV}~m_a^{-1})$. These initial conditions were chosen so that we can form a range of final structures; axion stars and black holes, like those found in \kcite{2017JCAP...03..055H,Michel:2018nzt}. We fix our boundary conditions to be periodic. 

We used varying adaptive mesh refinement (AMR) conditions based on the length scale, $\tilde{L}$, of the axion wave (discussed in \kap{subsect:AMR} in more detail). Each AMR level had a refinement ratio of two, with the coarsest grid set at a resolution of $64^3$. Convergence and stability of the simulations is discussed in \kap{subsect:convergence}.

\kap{subsect:constructing_initial_data} has a complete discussion on the initial conditions used. To summarise we calculate the initial value of trace of the extrinsic curvature, $K$, using 
\begin{equation}
K = -\sqrt{24 \pi G \langle \rho \rangle} \, ,
\end{equation}
where $ \langle \rho \rangle$ average initial energy density. The conformal factor, $\chi$, was calculated using a relaxation procedure until we reached a relative Hamiltonian constraint violation, $\mathcal{H}$
\begin{equation}
\mathcal{H} \equiv \frac{H_{center}}{16 \pi G \rho_{center}}  \, ,
\end{equation}
of $\mathcal{O}(0.1\%)$\footnote{During the relaxation routine the value of $H$ and $\rho$ at the centre of the simulation was also the max value of those variables}. The larger the length scale of the axion waves, the more numerically expensive the simulations were to perform due to an increase in the time scale of collapse, and a need for more refinement layers to track formation and the evolution of the resulting structure.

Blackhole formation is identified using a spherical horizon finder and the formation of an axion star was identified using an ``axion star location" script, which is detailed in \kap{subsect:ASF}.

\subsection{Axion Star Formation and Evolution}

Our simulations with initial conditions $M \leq$ $1.34$ $M_{\odot} \, 10^{-10} \mathrm{eV}~m_a^{-1}$ resulted in axion star formation (see \kfig{fig:moneyplot}). To compute the mass, $M_\star$, of the axion stars, we use \keq{eqn:mass_integral}, with the radius $R$ computed to be such that $\rho(R) = 0.05\rho_{max}$ -- this is a good approximation since the axion star sphericalizes rapidly.

\begin{figure*}
\centering
\includegraphics[width=1\columnwidth]{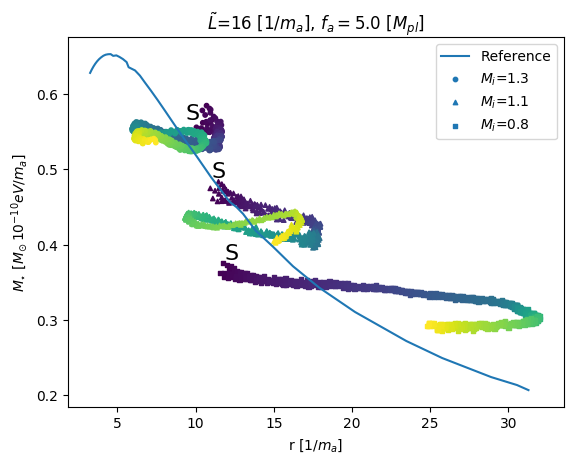}
\includegraphics[width=1\columnwidth]{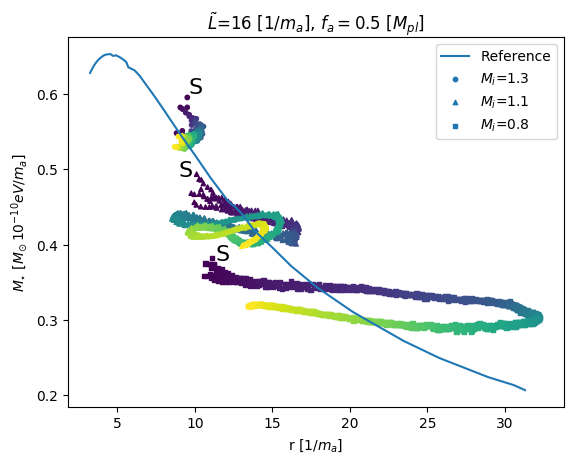}
\includegraphics[width=1\columnwidth]{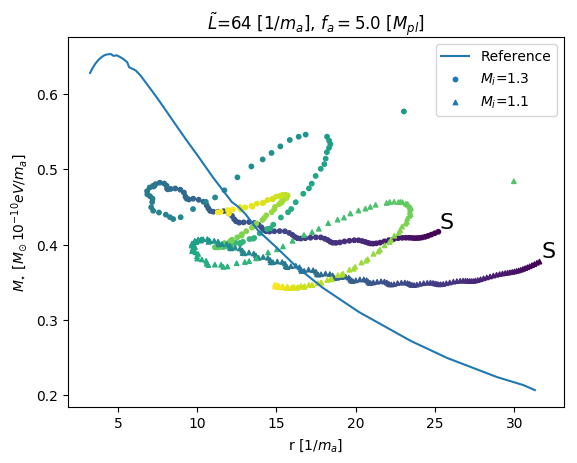}
\includegraphics[width=1\columnwidth]{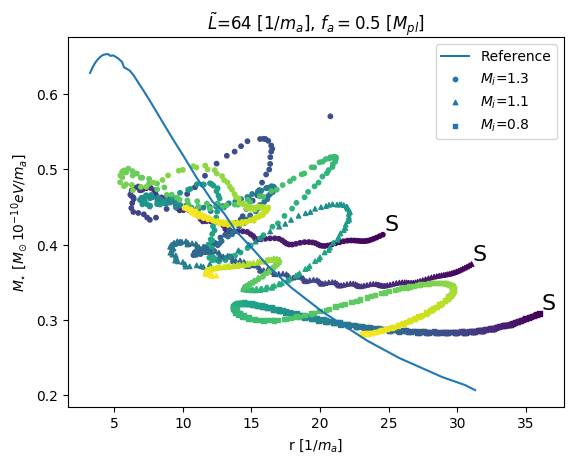}
\includegraphics[width=1\columnwidth]{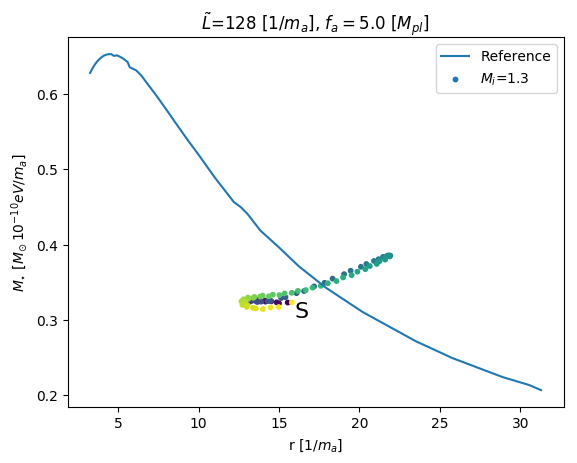}
\caption{Evolution of the mass-radius relation of all simulations whose end state was identified as an axion star. The reference line in the plots is the mass radius relation for an unexcited axion star, and the points on top indicate the evolution of the observed star forming. The evolution in time of the mass-radius relation is indicated by the colour of the point, with the darkest points being the earliest in the evolution and the lightest points being the end of the evolution. Additionally the start point of the evolution is indicated by an `S'. The mass radius relation fluctuates significantly over time, varying in a decaying way around the unexcited star value. This process is attributed to the formed stars having radial perturbative modes. Outliers on these graphs are due to the axion star finder, outlined in \kap{subsect:ASF} }
\label{fig:radial_modes}
\end{figure*}

\begin{figure}
\includegraphics[width=1\columnwidth]{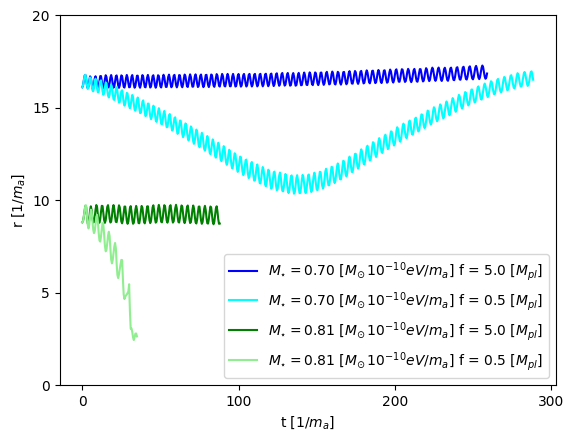}
\caption{Variation of the axion star radius over time for spherically symmetric initial data~\kcite{2017JCAP...03..055H,Helfer:2018vtq}. For $f_a= 5.0 \, M_{pl}$ the radial variation is negligible for both masses shown. When lowering $f_a$ to $0.5 \, M_{pl}$ it can be seen that the more massive axion star collapses to a blackhole, however for the less massive axion star a radial variation with a period of $300 \, m_a^{-1}$ develops. The radial variation shown here has a longer period compared to the most massive case for $\tilde{L}=16$, and shorter compared to $\tilde{L}=64$.}
\label{fig:static_axion}
\end{figure}

During the course of the evolution the axion stars were found to be stable (i.e. they do not disperse nor collapse into a BH), sphericalizing rapidly leaving only a dominant radially perturbative mode (see \kfig{fig:radial_modes}). \kfig{fig:static_axion} shows the variation of the radius of axion stars over time generated from spherically symmetric initial data \cite{2017JCAP...03..055H,Helfer:2018vtq}. The radial variation in the spherically symmetric case is long lived, and the computational cost of evolving the stars to their final end-state (presumably an unexcited star) is prohibitive. For $f_a= 5.0 \, M_{pl}$ the radial variation presented in \kfig{fig:radial_modes} is negligible for both masses shown. When lowering $f_a$ to $0.5 \, M_{pl}$ it can be seen that the more massive axion star collapses to a blackhole, however for the less massive axion star a radial variation with a period of $300 \, m_a^{-1}$ develops. The radial variation shown here has a longer period compared to the most massive case for $\tilde{L}=16$, and shorter compared to $\tilde{L}=64$. We conclude that the variation in radius of the stars from our formation process comes dominantly from decaying radially perturbative modes.

As has been shown in \kcite{Alcubierre:2003sx}, ground state axion stars span a family parameterized by the compactness parameter ${\cal C}$. When studying the compactness of axion stars formed by our collapse process vs this family, it can be seen that these formed stars oscillate around this family, hence represents stable stars. This can be seen in Fig.~\ref{fig:radial_modes}.

Finally, we compute the efficiency of the axion star formation process, which is defined as
\begin{equation}
\mathrm{Efficiency} \equiv \frac{\mathrm{Total~Initial~Mass}}{\mathrm{Mass~captured~in~AS}} \, .
\end{equation}
This is measured to range from $0.5$ for $\tilde{L}=128$ to $0.8$ for $\tilde{L}=16$. In other words, a large fraction of the initial mass forms the axion star. Since our simulation domain is periodic, and hence ``free scalar field energy'' has no place to disperse, we might worry that this may be due to a significant reabsorption. We observed that the axion star formed in $\mathcal{O}(10)$ ``box crossing" times and if reabsorption of the scalar field was big, we should see a modulation in $\phi$ at 10 times that frequency, which we do not. Hence we surmise that reabsorption is small and  expect that while in a dispersive environment the efficiency will be lower, it will not be significantly lower. 

\subsection{Black Hole Formation}

Meanwhile,  our simulations show that initial conditions with $M$ $=$ $2.14$ $M_{\odot} \, 10^{-10} \mathrm{eV}~m_a^{-1}$ resulted in black hole formation (see \kfig{fig:moneyplot}). This is consistent with the ``phase diagram'' presented in \kcite{2017JCAP...03..055H,Michel:2018nzt}. Similar to the axion star formation process, we found that the efficiency of black hole formation was $\mathcal{O}(1)$. 

\subsection{Dispersion Regime}
\label{sec:dispersion}
  
As shown in the phase diagram constructed in \kcite{2017JCAP...03..055H,Michel:2018nzt}, there exists ``dispersal regions'' where the axion star is not stable and disperses into scalar radiation. This occurs in regions with sufficiently low  $f_a$ and $M$. Due to the periodic domain, dispersed scalar fields will eventually fall back into a (possibly dispersing axion star), and hence we cannot probe this possibility.  Instead we use the phase diagram constructed in \kcite{2017JCAP...03..055H,Michel:2018nzt} for the analysis that follows.

\section{Axion Stars and Gravitational Waves}

Relativistic axion stars with high compactness can emit sufficiently strong GW signals which makes them possible targets of gravitational wave detectors. In this section, we will explore this possibility.

The phase diagram of Ref.~\cite{2017JCAP...03..055H} suggests that, for each value of $f_a$ below the ``triple point'', $f_{\rm TP}\approx 0.2 M_{pl}$, there are three phases of axion star: low mass stars are stable; those above a first critical mass, $M_{\rm disp.}$ are unstable to emission of relativistic axion waves; those above a second critical mass, $M_{\rm BH}$, collapse to BHs. Above the triple point, the dispersal phase no longer exists, and compact objects form for all masses. Our numerical simulations have verified that this same picture applies to the ``cosmological'' initial conditions of \keq{eqn:initial_phi}.

\subsection{Cosmological Formation of Axion Stars}

Axion stars are the (quasi-)stable end point of gravitational collapse of the axion field, and in simulations of dark matter structure formation have been observed to form in diverse conditions \cite{Levkov:2018kau, 2014NatPh..10..496S,PhysRevD.94.043513}

Non-relativistic axions stars are observed to form in simulations with coherent initial conditions, where they condense via monolithic collapse as the first generation of axion DM halos, with a population expected to inhabit the centres of all halos \cite{2014NatPh..10..496S,Schive:2014hza,2016PhRvD..94d3513S}. These axion stars form from the small  ($\zeta\approx 10^{-5}$) amplitude adiabatic curvature fluctuations which dominate the Universe on large scales, with the coherent initial conditions provided if Peccei-Quinn symmetry is broken during inflation. Cosmological simulations of this formation mechanism have only been performed for ultralight axions with $m_a\approx 10^{-22}\text{ eV}$. The corresponding axion stars in dwarf galaxies are too heavy to be relevant for LIGO. The formation mechanism, however, is expected to be operative for all axion masses in all dark matter halos~\cite{Du:2016aik}, potentially leading to relativistic cores in some region of parameter space.

Recently, axions stars were also shown to condense from highly incoherent initial conditions \cite{Levkov:2018kau}. This mechanism is expected to be active in axion ``miniclusters'' \cite{Sakharov,1988PhLB..205..228H,Kolb:1993hw,KHLOPOV1999105}, and indeed throughout any axion dark matter halo, potentially leading to spontaneous axion star formation. \kcite{Levkov:2018kau} proposes a growth rate that could make these axion stars reach relativistic masses if it does not quench. Mergers of such axion stars could also lead to mass increase.

Non-relativistic simulations like these provide realisations of the axion field with axion star locations. Dense peaks of this field will require individual relativistic simulations, and the evolution should resemble the cases that we have studied in the present work. In this model, the axion star population builds up over time in an astrophysical way, just as ordinary stars and BH do.

It is also possible that relativistic axion stars could form directly in the early Universe from large amplitude primordial fluctuations, a possibility we discuss in more detail in Section~\ref{sec:discussion}. In this case also, dense peaks of the axion field will evolve to relativistic axion stars as studied above. In this model, the axion star population resembles primordial BH.

\subsection{Peak Statistics}
\label{sec:statistics}

\begin{figure}
\center
\includegraphics[width=1\columnwidth]{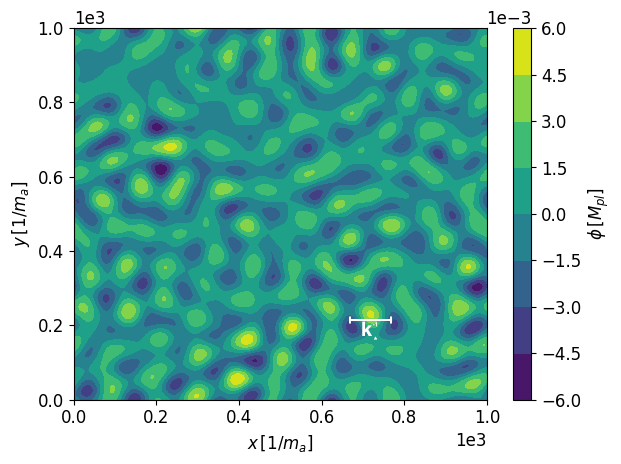}
\caption{A toy model realisation of the axion density field that can be filtered to locate candidate compact axion stars.}
\label{fig:xpower}
\end{figure}

Simulating a cosmological volume of initial conditions for the axion field with numerical relativity is not feasible. Instead we consider our simulations as representing isolated peaks in the density field. 

We consider a toy model for an axion density field containing large amplitude peaks that can be described by our numerical simulations. A simple mechanism to form massive, dense, primordial AS in the LIGO band is to enhance the axion power spectrum by a Gaussian bump on small scales:
\begin{equation} 
\label{eqn:power_spectrum} 
P_{\delta\phi} \equiv  A \exp \left[\frac{-(k-k_\star)^2}{2\sigma_k^2}\right]\, , 
\end{equation} 
where $A$ is the amplitude, $k_\star$ the central mode, and $\sigma_k\ll k_\star$ the width. The amplitude has units $M_{pl}^2 m_a^{-3}$. \kfig{fig:xpower} shows the toy model for the axion density field with $k_*^{-1}$ roughly equal to the size of a peak described by our numerical simulations. For any axion density field like this toy example (e.g. density field and axion star location inside a DM halo), we can calculate the mass distribution of axion stars and BH formed by the extreme peaks by analogy to the theory of critical collapse for BHs, and to the Press-Schechter theory of cosmological structure formation. 

\begin{figure*}
\center
\includegraphics[width=1\columnwidth]{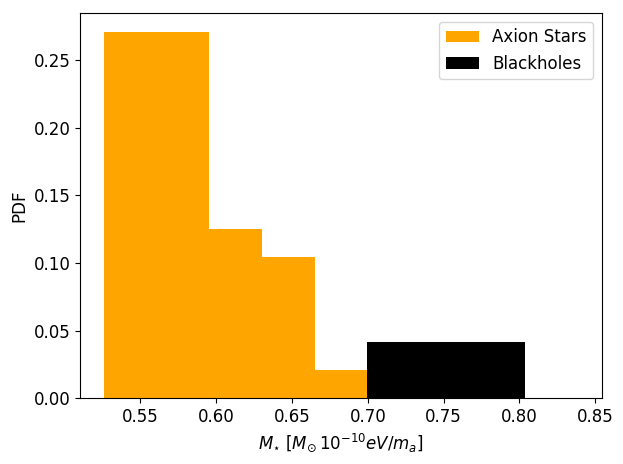}
\includegraphics[width=1\columnwidth]{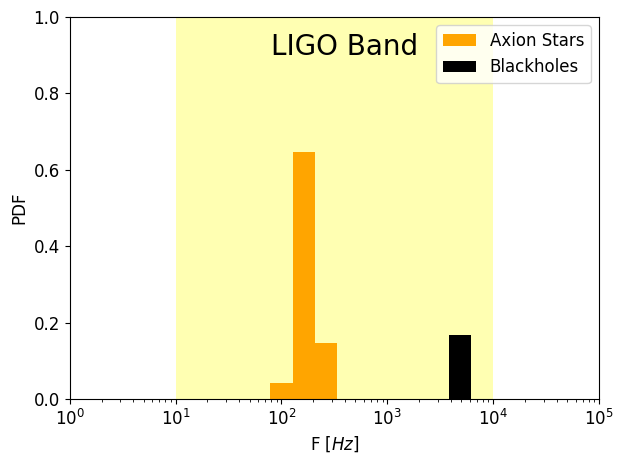}
\caption{Left panel: Peaks of the toy model density field are assigned masses as axion stars and black holes according to the location on the phase diagram with $f_a= 0.5$ $M_{pl}$. Right panel: Using \keq{eqn:fcomp} and the $\mathcal{C}(M)$ relation for axion stars, we calculate the gravitational wave frequencies for axion star and BH equal mass binary mergers. }
\label{fig:objects}
\end{figure*}

\begin{figure*}
\center
\includegraphics[width=1\columnwidth]{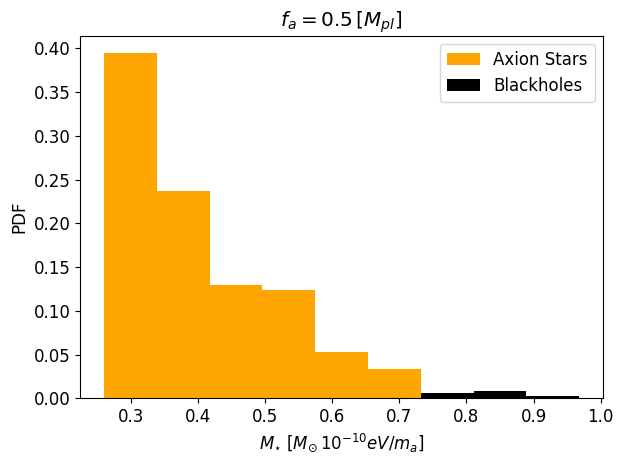}
\includegraphics[width=1\columnwidth]{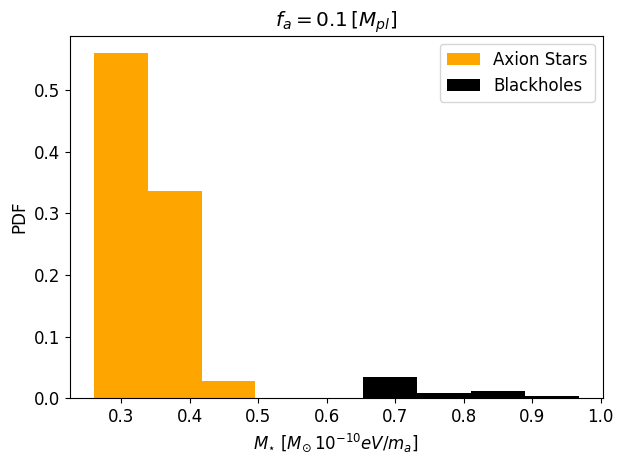}
\caption{Peaks of the toy model density as \kfig{fig:objects} with $f_a= 0.5$ $M_{pl}$ (left panel) and $f_a= 0.1$ $M_{pl}$ (right panel). Note that as $f_a$ dips below the triple point \kcite{2017JCAP...03..055H}, a dispersal gap appears between the formation of black holes and axion stars.} 
\label{fig:mass_gap}
\end{figure*}

The peak statistics are determined by thresholding the field, and are classified using a two pass connected component clustering algorithm, see \kap{app:CCL}. We partition the distribution for $\tilde{M}_{\star}(f_a)$ according to the critical masses $M_{\rm disp.}$ and $M_{\rm BH}$ in the phase diagram~\cite{2017JCAP...03..055H}. In our numerical simulations, due to the construction with periodic boundary conditions and an isolated fluctuation, fluctuations of all amplitudes leading to axion stars with radius smaller than the box size will eventually collapse under self-gravity. Thus we cannot determine the critical threshold for axion star collapse. However, we are only interested phenomenologically in the densest, and thus most massive stars, and so we threshold our field for a minimum compactness of the final axion star assuming that at least these most compact objects successfully collapse.

\kfig{fig:objects} and \kfig{fig:mass_gap} apply such a thresholding and labelling to a statistically representative realisation of the toy model field of \kfig{fig:xpower}, taking only the one percent densest peaks. The labelled peaks span from axion stars, to a mix of axion stars and blackholes, to primarily blackholes depending on the field variance. Lowering $f_a$ below $f_{\rm TP}$ results in the formation of a mass gap of objects. In particular, as $f_a$ dips below the triple point at $f_{\rm TP}=0.2 M_{pl}$, a mass gap appears between the masses of formed axion stars and black holes. This mass gap is a characteristic feature of axion stars, hence the observation of a mass gap in the power spectrum of compact objects is a clear method of identifying the existence of axions in the universe.

We can also estimate the frequency of gravitational waves emitted by an axion star-axion star or BH-BH binary merger using Eq.~\eqref{eqn:fcomp}. Assuming binary mergers from the density field simply takes the field statistics as representative: no merger rate is calculated. \kfig{fig:objects} shows the frequencies with an axion mass of $m_a \approx 10^{-10} \mathrm{eV}$. We observe that, for this distribution of peaks, axion star binary coalescence, as well as BH-BH binary coalescence from collapsed axion stars fall in the LIGO band.

\section{Discussion}
\label{sec:discussion}

It is important to note that only a small fraction of the DM density should be contained in axion stars and primordial BH. It is still an intriguing possibility that some GW events detectable by LIGO might be due to primordial BH, and the the distribution of GW events could be used to confirm this~\cite{Bird:2016dcv}. Similarly, GW events and their distribution could confirm the existence of a fraction of DM in axion stars and BH formed from their collapse \cite{Croon:2018ftb,Brito:2017zvb}.

Recent work with scalar compact objects head on mergers \cite{Cardoso:2016oxy,Palenzuela:2006wp,Helfer:2018vtq,Choptuik:2009ww} indicates distinctions in the gravitational wave signal with respect to blackholes. If these distinctions also exist in binary coalescence (see \cite{Sennett:2017etc,Palenzuela:2017kcg,Bezares:2017mzk} for boson star inspirals), a single GW event could be a smoking gun for the existence of axion stars. The end state mass spectrum from a number of such events could be used to determine the axion decay constant. 

One possible mechanism to form compact axions stars in the early Universe is to enhance the axion power spectrum, $\langle\delta\phi(k)\delta\phi(k')\rangle$, on small scales, similarly to production methods for primordial black holes (see Ref.~\cite{Carr:2005bd} and references therein). Such enhanced axion fluctuations on small scales arise generically in models of inflation in which the radial mode of the Peccei-Quinn field evolves as a spectator, leading to a strongly blue axion spectrum~\cite{Kasuya:2009up,Chung:2016wvv}.

Unfortunately, under standard cosmological assumptions such a power spectrum cannot form compact axion stars. A fluctuation in the axion field at early times is isocurvature (since the axions are subdominant compared to the radiation). During the radiation epoch, the linear transfer function of isocurvature overdensities is close to unity~\cite{1986MNRAS.218..103E}. This implies that, between the time the axion field becomes non-relativistic, $H(a_{\rm osc})\approx m_a$, and matter-radiation equality, $a_{\rm eq}$, the field fluctuation is redshifted as $\delta\phi\sim a^{-3/2}$. For the axion masses of interest, $m_a\approx 10^{-10}\text{ eV}\Rightarrow a_{\rm osc}/a_{\rm eq}\approx 10^{-14}$, giving huge redshift factors. 

Collapse of primordial fluctuations could occur during a putative early matter dominated phase~\cite{Khlopov:1980mg,2016ApJ...833...61H} (as expected in supersymmetric models, e.g. Refs.~\cite{1983PhLB..131...59C,2014PhRvD..89j3513I}), or during reheating if the equation of state is in the correct regime. Study of collapse of axion stars during such a period, or from primordial curvature perturbations in the radiation era, requires additional simulations that account for the background fluid in addition to the axion scalar field. This interesting possibility, incorporating fluids into \textsc{GRChombo}, will be the subject of future study.

An axion star in the LIGO band requires $m_a\approx 10^{-10}\text{ eV}$. The QCD axion with this mass has $f_a\approx 10^{16}\text{ GeV}$, which could possibly be detected directly by ABRACADABRA~\cite{Kahn:2016aff} or CASPEr~\cite{2014PhRvX...4b1030B}. However, with this low value of $f_a$ axion stars cannot reach the required high compactness before becoming unstable. This leads to the interesting conclusion that any future observation of GWs from axion stars would imply the existence of contributions to the axion spectrum beyond QCD, and could thus lend support to the idea of a ``String Axiverse''~\cite{Arvanitaki:2009fg} or other non-standard axion scenarios~\cite{DiLuzio:2016sbl}.

\paragraph*{Acknowledgments}  

We acknowledge useful conversations with Mustafa Amin, Malcolm Fairbairn, Matthew McCullough, Jens Niemeyer and Helvi Witek. We would also like to thank the GRChombo team (http://www.grchombo.org/) and the COSMOS team at DAMTP, Cambridge University for their ongoing technical support. DJEMs research at the University of G\"{o}ttingen is funded by the Alexander von Humboldt Foundation and the German Federal Ministry of Education and Research. EL is supported by STFC AGP grant ST/P000606/1, and JW is supported by a STFC PhD studentship.  Numerical simulations were performed on the COSMOS supercomputer, funded by DIRAC/BIS and on BSC Marenostrum IV via PRACE grant Tier-0 PPFPWG, by the Supercomputing Centre of Galicia and La Palma Astrophysics Centre via BSC/RES grants AECT-2017-2-0011 and AECT-2017-3-0009 and on SurfSara Cartesius under Tier-1 PRACE grant DECI-14 14DECI0017. Some simulation results were analyzed using the visualization toolkit YT \cite{2011ApJS..192....9T}. Matplotlib \cite{Hunter:2007} was used to generate the plots seen throughout the paper, and Numpy \cite{scipy} was used for the statistical analysis.

\bibliographystyle{h-physrev3.bst}

\bibliography{paper}
\newpage
\appendix

\section{Two Pass Connected Component Labelling}
\label{app:CCL}
The connected component labelling (CCL) procedure assigns a unique label to a set of connected target pixels in a binary image \cite{10.1007/978-3-642-25330-0_40}. We can construct a binary image from an 2D array, in our case the energy density of the $\phi$ field generated by our toy power spectrum, by assigning a 0 to all elements in the array that are below a cutoff threshold, and a 1 to all those that are above. A subset of the binary image is called connected if for any two points $P$ and $Q$ of the subset there exists a sequence of points $P= P_0, P_1, P_2 ..., P_{n-1}, P_n=Q$ such that $P_i$ is a neighbour of $P_{i-1}$ \cite{Rosenfeld:1966:SOD:321356.321357}. The definition of a connection relies on that of a pixel's neighbourhood, if this includes 4 neighbours it is said to be 4-connected, and if it includes 8 neighbours it is said to be 8-connected \kcite{797615}. 

We use a specific group of CCL, known as two-pass algorithms, to label peaks. Two pass complete labelling in two scans: during the first scan they assign provisional labels to pixels and record label equivalencies. Label equivalencies are the resolved during or after the the first scan. During the second scan, all equivalent labels are replaced by their representative label \cite{HE20091977,Suzuki:2003:LCL:780779.780781}. We use a two-pass algorithm that use 4-connected to define a connection. Algorithmically, we did the following:

\begin{algorithmic}
\Loop { p}
\If { p $>0$}
    \If {p above !=0 and p left =0}
    \State { pl = pl above}
    \ElsIf { p above =0 and p left !=0}
    \State { pl = pl left}
    \ElsIf { p above !=0 and p left !=0}
    \State { pl = min(pl above, pl left)}
    \State { record pl dependancy}
    \Else 
    \State { pl = new label}
    \EndIf
\EndIf
\EndLoop
\end{algorithmic}
where p is a pixel, pl is a pixel label, and min is a function that chooses the minimum of two values. The label equivalencies were then processed such that consecutive labels where generated, and then a second pass would the replace all equivalent labels. This algorithm provides a description of how the peaks where labelled. The Numpy \cite{scipy} CCL algorithm was used for the analysis presented in this paper.

\section{GRChombo code}
\label{sect:GRChombo}

This appendix summarises the key features of the numerical relativity code $\textsc{GRChombo}$. For a more full discussion see \kcite{Clough:2015sqa}, and the $\textsc{GRChombo}$ website at \url{http://grchombo.org}, which includes links to movies of simulations using the code.

\subsection{Numerical implementation}
$\textsc{GRChombo}$ is a multi-purpose numerical relativity code, which is built on top of the open source $\mathtt{Chombo}$ framework. $\mathtt{Chombo}$ is a set of tools developed by Lawrence Berkeley National Laboratory for implementing block-structured AMR in order to solve partial differential equations \cite{Chombo}.

The key features of $\mathtt{Chombo}$  are:
\begin{itemize}
\item{\emph{C++ class structure}: $\mathtt{Chombo}$ is primarily written in the C++ language, using the class structure inherent in that language to separate the various evolution and update processes.}
\item{\emph{Adaptive Mesh Refinement}: $\mathtt{Chombo}$ provides Berger-Oliger style \cite{bergeroliger,BergerColella} AMR with Berger-Rigoutsos \cite{BergerRigoutsis91} block-structured grid generation. Chombo supports full non-trivial mesh topology -- i.e. many-boxes-in-many-boxes. The user is required to specify regridding criteria, which is usually based on setting a maximum threshold for the change in a variable across a gridpoint.}
\item{\emph{MPI scalability}: $\mathtt{Chombo}$ contains parallel infrastructure which gives it the ability to scale efficiently to several thousand CPU-cores per run. It uses an inbuilt load balancing algorithm, with Morton ordering to map grid responsibility to neighbouring processors in order to optimize processor number scaling.}
\item{\emph{Standardized Output and Visualization}: $\mathtt{Chombo}$ uses the $\mathtt{HDF5}$ output format, which is supported by many popular visualization tools such as $\mathtt{VisIt}$. In addition, the output files can be used as input files if one chooses to continue a previously stopped run -- i.e. the output files are also checkpoint files.}
\end{itemize}

The key features of $\textsc{GRChombo}$ are:
\begin{itemize}
\item{\emph{BSSN formalism with moving puncture}: $\textsc{GRChombo}$ evolves the Einstein equation in the BSSN formalism with scalar matter. Singularities of black holes are managed using the moving puncture gauge conditions \cite{Campanelli:2005dd, Baker:2005vv}. These evolution equations and gauge conditions are detailed further below. There is an option to turn on CCZ4 constraint damping terms if required, but this was not used in this work.}
\item{\emph{4th order discretisation in space and time}: We use the method of lines with 4th order spatial stencils and a 4th order Runge-Kutta time update. We use symmetric stencils for spatial derivatives, except for the advection derivatives (of the form $\beta^i \partial_i F$) for which we use one-sided/upwinded stencils. In \kcite{Clough:2015sqa} it was shown that the convergence is approximately 4th order without regridding, but reduces to 3rd order convergence with regridding effects.}
\item{\emph{Kreiss-Oliger dissipation}: Kreiss-Oliger dissipation is used to control errors, from both truncation and the interpolation associated with regridding.}
\item{\emph{Boundary conditions}: We use either periodic boundaries or Sommerfeld boundary conditions \cite{Alcubierre:2002kk}, which allow outgoing waves to exit the grid with minimal reflections. For many simulations, the AMR ability allows us to set the boundaries far enough away so that reflections do not affect the results during simulation time. In this work only periodic boundary conditions were used.}
\item{\emph{Initial Conditions}: In principle any initial conditions can be used, for example, where solutions to the constraints have been found numerically, these can be read into the grid using a simple first order interpolation. Note that $\textsc{GRChombo}$ itself does not currently solve the constraints for the initial conditions, although it can be used to relax the Hamiltonian constraint for the value of the conformal factor $\chi$ where the other variables are assumed to solve the momentum constraint and admit solutions consistent with the boundary conditions.}
\item{\emph{Diagnostics}: $\textsc{GRChombo}$ permits the user to monitor the Hamiltonian and momentum constraint violation, find spherically symmetric apparent horizons, extract gravitational waves and calculate ADM mass and momenta values.}
\end{itemize}

\subsection{Gauge choice}

$\textsc{GRChombo}$ uses the BSSN formalism \cite{Baumgarte:1998te,Nakamura,Shibata:1995we} of the Einstein equation in 3+1 dimensions. This is similar to the more well known ADM decomposition \cite{PhysRev.116.1322}, but is more stable numerically. The 4 dimensional spacetime metric is decomposed into a spatial metric on a 3 dimensional spatial hypersurface, $\gamma_{ij}$, and an extrinsic curvature $K_{ij}$, which are both evolved along a chosen local time coordinate $t$. Since one is free to choose what is space and what is time, the gauge choice must also be specified.
The line element of the decomposition is
\begin{equation}
ds^2=-\alpha^2\,dt^2+\gamma_{ij}(dx^i + \beta^i\,dt)(dx^j + \beta^j\,dt)\,,
\end{equation}
where $\alpha$ and $\beta^i$ are the lapse and shift, the gauge parameters.  These parameters are specified on the initial hypersurface (see below) and then allowed to evolve using gauge-driver equations, in accordance with the puncture gauge \cite{Campanelli:2005dd,Baker:2005vv}, for which the evolution equations are
\begin{align} \label{eqn:MovingPuncture}
&\partial_t \alpha = - \mu \alpha K + \beta^i \partial_i \alpha ~ , \\
&\partial_t \beta^i = B^i ~ , \\
&\partial_t B^i = \frac{3}{4} \partial_t \Gamma^i - \eta B^i ~ ,
\end{align}
where the constants $\eta$, of order $1/M_{ADM}$, and $\mu$, of order 1, may be varied by the user to improve stability. The effect of the moving puncture gauge is to avoid resolving the central singularity of any black hole that may form. It was shown that in this gauge the central gridpoints asymptote to a fixed radius within the event horizon, the so-called ``trumpet'' solution described in \kcite{Hannam:2008sg}. Thus explicit numerical excision of the central singularity is not required. While constraint violation may occur at the central point due to taking gradients across the puncture, these remain within the horizon and do not propagate into the outside spacetime.

\subsection{Evolution equations}

In $\textsc{GRChombo}$ the induced metric is decomposed as
\begin{equation}
\gamma_{ij}=\frac{1}{\chi}\,\tilde\gamma_{ij} \quad \det\tilde\gamma_{ij}=1 \quad \chi = \left(\det\gamma_{ij}\right)^{-\frac{1}{3}}  ~ .
\end{equation}
The extrinsic curvature is decomposed into its trace, $K=\gamma^{ij}\,K_{ij}$, and its traceless part $\tilde\gamma^{ij}\,\tilde A_{ij}=0$ as
\begin{equation}
K_{ij}=\frac{1}{\chi}\left(\tilde A_{ij} + \frac{1}{3}\,K\,\tilde\gamma_{ij}\right) ~ .
\end{equation}
The conformal connections $\tilde\Gamma^i=\tilde\gamma^{jk}\,\tilde\Gamma^i_{~jk}$ where $\tilde\Gamma^i_{~jk}$ are the Christoffel symbols associated with the conformal metric $\tilde\gamma_{ij}$. The evolution equations for BSSN are then 

\begin{align}
&\partial_t\chi=\frac{2}{3}\,\alpha\,\chi\, K - \frac{2}{3}\,\chi \,\partial_k \beta^k + \beta^k\,\partial_k \chi ~ , \label{eqn:dtchi2} \\
&\partial_t\tilde\gamma_{ij} =-2\,\alpha\, \tilde A_{ij}+\tilde \gamma_{ik}\,\partial_j\beta^k+\tilde \gamma_{jk}\,\partial_i\beta^k \nonumber \\
&\hspace{1.3cm} -\frac{2}{3}\,\tilde \gamma_{ij}\,\partial_k\beta^k +\beta^k\,\partial_k \tilde \gamma_{ij} ~ , \label{eqn:dttgamma2} \\
&\partial_t K = -\gamma^{ij}D_i D_j \alpha + \alpha\left(\tilde{A}_{ij} \tilde{A}^{ij} + \frac{1}{3} K^2 \right) \nonumber \\
&\hspace{1.3cm} + \beta^i\partial_iK + 4\pi G\,\alpha (\rho + S) \label{eqn:dtK2} ~ , \\
&\partial_t\tilde A_{ij} = \left[-D_iD_j \alpha + \chi \alpha\left( R_{ij} - 8\pi G\,\alpha \,S_{ij}\right)\right]^\textrm{TF} \nonumber \\
&\hspace{1.3cm} + \alpha (K \tilde A_{ij} - 2 \tilde A_{il}\,\tilde A^l{}_j)  \nonumber \\
&\hspace{1.3cm} + \tilde A_{ik}\, \partial_j\beta^k + \tilde A_{jk}\,\partial_i\beta^k \nonumber \\
&\hspace{1.3cm} -\frac{2}{3}\,\tilde A_{ij}\,\partial_k\beta^k+\beta^k\,\partial_k \tilde A_{ij}\,   \label{eqn:dtAij2} ~, \\
&\partial_t \tilde \Gamma^i=2\,\alpha\left(\tilde\Gamma^i_{jk}\,\tilde A^{jk}-\frac{2}{3}\,\tilde\gamma^{ij}\partial_j K - \frac{3}{2}\,\tilde A^{ij}\frac{\partial_j \chi}{\chi}\right) \nonumber \\
&\hspace{1.3cm} -2\,\tilde A^{ij}\,\partial_j \alpha +\beta^k\partial_k \tilde\Gamma^{i} \nonumber \\
&\hspace{1.3cm} +\tilde\gamma^{jk}\partial_j\partial_k \beta^i +\frac{1}{3}\,\tilde\gamma^{ij}\partial_j \partial_k\beta^k \nonumber \\
&\hspace{1.3cm} + \frac{2}{3}\,\tilde\Gamma^i\,\partial_k \beta^k -\tilde\Gamma^k\partial_k \beta^i - 16\pi G,\alpha\,\tilde\gamma^{ij}\,S_j ~ . \label{eqn:dtgamma2}
\end{align}
The scalar field matter evolution equations are
\begin{align}
&\partial_t \phi = \alpha \Pi_M +\beta^i\partial_i \phi \label{eqn:dtphi2} ~ , \\
&\partial_t \Pi_M=\beta^i\partial_i \Pi_M + \alpha\partial_i\partial^i \phi + \partial_i \phi\partial^i \alpha \\
&\hspace{1.3cm} +\alpha\left(K\Pi_M-\gamma^{ij}\Gamma^k_{ij}\partial_k \phi+\frac{dV}{d\phi}\right) \label{eqn:dtphiM2} ~ ,
\end{align}
where the second order Klein Gordon equation has been decomposed into two first order equations as is usual.

The stress energy tensor for a single scalar field is
\begin{equation}
T_{ab} = \nabla_a \phi \nabla_b \phi - \frac{1}{2} g_{ab} (\nabla_c \phi \, \nabla^c \phi + 2V) \ .
\end{equation}
and the various components of the matter stress tensor are calculated from this as
\begin{align}
&\rho = n_a\,n_b\,T^{ab}\,,\quad S_i = -\gamma_{ia}\,n_b\,T^{ab}\,, \nonumber \\
&S_{ij} = \gamma_{ia}\,\gamma_{jb}\,T^{ab}\,,\quad S = \gamma^{ij}\,S_{ij} \, .
\label{eq:Mattereqns}
\end{align}

\noindent The Hamiltonian constraint is
\begin{equation}
\mathcal{H} = R + K^2-K_{ij}K^{ij}-16G\pi \rho \, . \label{eqn:Ham}
\end{equation}
\noindent The momentum constraint is
\begin{equation}
\mathcal{M}_i = D^j (K_{ij} - \gamma_{ij} K) - 8\pi G S_i \, .  \label{eqn:Mom}
\end{equation}

\subsection{Constructing Initial data}
\label{subsect:constructing_initial_data}
We construct our initial data in the same was as in \kcite{Clough:2016ymm}, however the key details of the method are reproduced here for convenience. We choose $\alpha = 1$ and $\beta_i=0$ and hence on the initial hypersurface the initial gradient energy is
\begin{align}
& \rho_{grad} \equiv \frac{1}{2} \gamma^{ij} \partial_i \phi \partial_j \phi
\label{eqn:initial_gradient_energy}
\end{align}
We also introduce the notation for the kinetic term
\begin{align}
& \eta = \frac{1}{\alpha} \left ( \partial_t \phi - \beta^{k} \right )
\label{eqn:initial_kinetic_energy}
\end{align}
Which is zero on our initial hypersurface. Hence initially
\begin{align}
& \rho=  \frac{1}{2} \gamma^{ij} \partial_i \phi \partial_j \phi + V
\label{eqn:initial_rho}
\end{align}
Our constraint equations become
\begin{multline}
\tilde{D}^2\chi -\frac{5}{4 \chi}\tilde{\gamma}^{ij}\tilde{D}_i\chi \tilde{D}_j\chi \\ + \frac{\chi \tilde{R}}{2} + \frac{K^2}{3} - \frac{1}{2} \tilde{A}_{ij}\tilde{A}^{ij} = 8\pi G \rho \label{eqn:HamCon}\,,
\end{multline}
and
\begin{equation}
\tilde{D}_j \tilde{A}^{ij} -\frac{3}{2 \chi} \tilde{A}^{ij}\tilde{D}_j \chi -\frac{2}{3}\tilde{\gamma}^{ij}\tilde{D}_j K = 8\pi G \eta \tilde{\gamma}^{ij} \partial_j \phi\,. \label{eqn:MomCon}
\end{equation}
Next we want to specify the initial conditions for the metric $\gamma_{ij}$ and the extrinsic curvature $K_{ij}$. We can make the simplifying assumption that the metric is conformally flat and the traceless part of the extrinsic curvature $K_{ij}$ is zero everywhere on the initial hyperslice
\begin{equation}
\tilde{\gamma}_{ij} = \delta_{ij}\,,  \label{eqn:conformalcond}
\end{equation}
and
\begin{equation}
\tilde{A}_{ij} = 0\,. \label{eqn:tracelessK}
\end{equation}
We now need to specify the values of $K$ and $\chi$ on the initial hyperslice. \keq{eqn:MomCon} is trivially satisfied for $K=const$, however, in order to satisfy \keq{eqn:HamCon}, and the periodic boundary conditions or $\chi$, $K^2/24\pi$ needs to lie close to the average initial energy density for the hypersurface. Therefore for simplicity we choose it equal to the average initial energy density, approximating the metric to be Euclidean
\begin{equation}
K = -\sqrt{24 \pi G \langle \rho \rangle}\,, \label{eqn:Kavg}
\end{equation}
with
\begin{equation}
\label{eq:App:rho_initial}
\rho = \frac{1}{2}(\partial_i \phi)^2 + V(\phi)\,,
\end{equation}
where $\langle X \rangle = {\cal V}^{-1} \int X~d{\cal V}$ indicates the average over the spatial volume ${\cal V}$ of the quantity $X$.  Once $K$ is chosen, the initial field profile and the Hamiltonian constraint then fully determine the conformal factor $\chi$ (which we solve for using numerical relaxation).
\subsection{AMR Condition}
\label{subsect:AMR}
All simulations shared a coarsest grid of $64^3$. Locally, the expansion of our spacetime is roughly 
\begin{equation}
a \equiv \frac{1}{\sqrt{\chi}}.
\end{equation}
\begin{figure}
\includegraphics[width=1\linewidth]{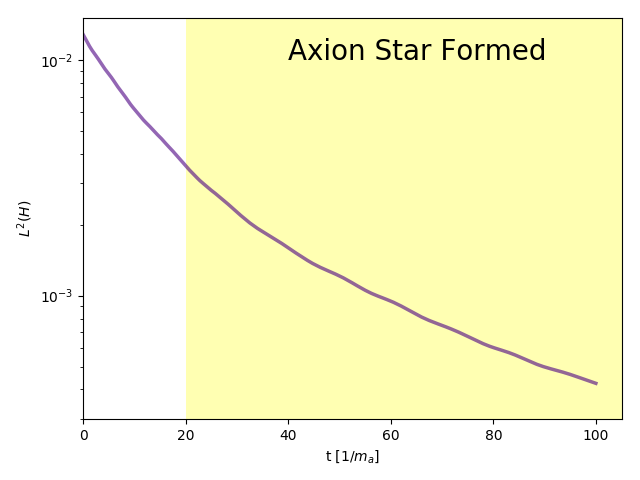}
\caption{The plot shows the $L^2$ norm \keq{eq:L2H} of the Hamiltonian constraint violation over time for a simulation that forms an axion star, with an initial total mass of $M= 1.34$ $M_{\odot} \, 10^{-10} \mathrm{eV}~m_a^{-1}$, $f_a=5.0$ $M_{pl}$ and $\tilde{L}= 16$ $m_a^{-1}$. The spikes in the plot are due to the regridding in the simulation and are rapidly damped.}
\label{fig_L2H}
\end{figure}
\begin{figure}
\includegraphics[width=1\linewidth]{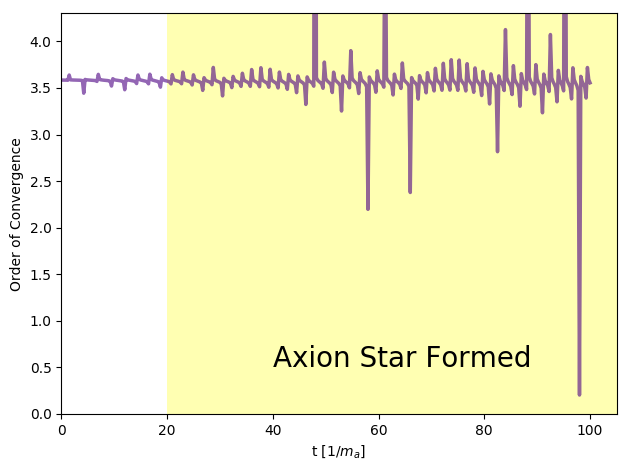}
\caption{Convergence test for $\phi_{center}$ showing a convergence between 3rd and 4th order. The convergence test is done with a fixed grid with three different resolutions of $0.25$ $m_a^{-1}$, $0.125$ $m_a^{-1}$ and $0.0625$ $m_a^{-1}$. Our evolution scheme is 4th order and the variation in the convergence is due to $\phi$ passing through $0$ during the evolution.}
\label{fig_convergence}
\end{figure}

Since the timescale of formation of objects varies, this means that the physical length scales of the problem do not necessarily track the grid, and hence requires a rescale regridding threshold. We set our threshold to be triggered by high gradients in $K$ and a scaled version of the gradients of $\rho$. These conditions track gravitational collapse in our simulations.  For $\tilde{L}=16$ we scaled our regridding of the gradients of $\rho$ as
\begin{equation}
\rho_t\frac{\nabla \rho}{\sqrt{\chi}} \, , \nonumber
\end{equation}
where $\rho_t$ is a numerical regridding threshold set at the beginning of the simulation. It was set to a value of $20\rho_{max}^{star} $, where $\rho_{max}^{star}$ is the maximum value of $\rho$ that a star of half total box mass would have. If half the total box mass was greater then the most stable axion star, then $\rho_{max}^{star}$ was set to be the for the highest stable axion star. For $\tilde{L}=64$ and $\tilde{L}=128$ we found that this condition was not enough for optimum regridding. Below an amount of layers (3 for 64, and 4 for 128), we added an additional regrid condition 
\begin{equation}
\rho_a\frac{\nabla \rho}{\chi^{\frac{3}{2}}} \, , \nonumber
\end{equation}
where $\rho_a$ is an additional regridding threshold. $\rho_a$ was chosen to be $10 \rho_{max}^{box}$, where $\rho_{max}^{box}$ is the maximum value of $\rho$ in the simulation at $t=0$. Once the correct thresholds where chosen, these regrid conditions would effectively follow the gravitational collapse in the simulations.
\subsection{Convergence and Stability}
\label{subsect:convergence}
We use the following to measure the volume averaged Hamiltonian constraint violation:
\begin{equation}\label{eq:L2H}
L^2(H) =\sqrt{\frac{1}{V} \int_V |\mathcal{H}^2| dV},
\end{equation}
where $V$ is the box volume with the interior of the apparent horizon excised. As can be seen in \kfig{fig_L2H}, we have good control over the constraint violation throughout the simulation.

We test the convergence of our simulations with the formation of an Axion Star with initial total mass of $M= 1.34$ $M_{\odot} \, 10^{-10} \mathrm{eV}~m_a^{-1}$, $f_a=5.0$ $M_{pl}$ and $\tilde{L}=16$ $m_a^{-1}$. We use a fixed grid for the convergence test with resolutions of $0.25$ $m_a^{-1}$, $0.125$ $m_a^{-1}$ and $0.0625$ $m_a^{-1}$. The results are shown in \kfig{fig_convergence}, where we obtain an order of convergence between 3rd and 4th order on average. The variation of the in the convergence test is due to the methodology, where we extract values of $\phi$ at the centre of the grid. $\phi$ passes through $0$ during the evolution, that causes the spikes present in the convergence test.

\subsection{Axion Star Location}

\label{subsect:ASF}
To confirm the that a resulting object was an axion star, and to track its subsequent evolution, an ``axion star finder" script was written, and ran in post-processing. The finder would look at the central density in the simulation, and locate the value and location of it's maximum, $\rho_{max}$. The radius in which the value of $\rho$ had dropped to $5\%$ of $\rho_{max}$ was calculated, and then the total mass was defined as the integrated density within a sphere of that radius. The radius of the object was adjusted for expansion.

In \kfig{fig:radial_modes}, there are some points that can be considered to be outliers. When the script looks for the maximum value and location of $\rho$, if at that point in time there are two maximum points in the central region, it causes the script to not return the true radius of the object, and hence the calculated mass will also not be correct. This was not a frequent occurrence, and the cause of it is easily confirmed.

\end{document}